\documentclass[american,polish,english,twocolumn,aps,prb,reprint,doi,longbibliography]{revtex4-2}
\usepackage[T1]{fontenc}
\usepackage[utf8]{inputenc}
\usepackage{geometry}
\usepackage{xcolor}
\usepackage{babel}
\usepackage{amsmath}
\usepackage{amssymb}
\usepackage{graphicx}
\usepackage{esint}
\PassOptionsToPackage{normalem}{ulem}
\usepackage{ulem}
\usepackage[pdfusetitle,
 bookmarks=true,bookmarksnumbered=false,bookmarksopen=false,
 breaklinks=false,pdfborder={0 0 1},backref=false,colorlinks=false]
 {hyperref}

\makeatletter

\newcommand{\lyxmathsym}[1]{\ifmmode\begingroup\def\b@ld{bold}
  \text{\ifx\math@version\b@ld\bfseries\fi#1}\endgroup\else#1\fi}

\providecolor{lyxadded}{rgb}{0,0,1}
\providecolor{lyxdeleted}{rgb}{1,0,0}
\DeclareRobustCommand{\mklyxadded}[1]{\bgroup\color{lyxadded}{}#1\egroup}
\DeclareRobustCommand{\mklyxdeleted}[1]{\bgroup\color{lyxdeleted}\mklyxsout{#1}\egroup}
\DeclareRobustCommand{\mklyxsout}[1]{\ifx\\#1\else\sout{#1}\fi}

\DeclareRobustCommand{\lyxdeleted}[4][]{\texorpdfstring{\mklyxdeleted{#4}}{}}

\makeatother

\begin{document}
\title{Finite-Temperature Quantum-Rotor Approach for Ultracold Bosons in
Optical Lattices}
\author{M. Rodríguez Martín and T. A. Zaleski}
\address{Institute of Low Temperature and Structure Research, PAS, Okólna 2,
50-422 Wroclaw, Poland}
\begin{abstract}
Interacting bosons in optical lattices directly expose quantum phases
in a clean, highly controllable environment. This requires engineering
systems with very low entropies, but the resulting temperature--interaction
ratios $T/U$ of present experiments remain well above the domain
where zero-temperature theories are expected to be reliable. The quantum-rotor
approach (QRA), while analytically powerful and extremely flexible,
inherits ground-state phase correlations and therefore breaks down
once thermal winding of the phase field becomes significant. Here
we construct a finite-temperature extension of QRA by (i) performing
resummation of winding-number contributions for temperatures $k_{B}T/U\lesssim0.2$
and (ii) developing an auxiliary-variable expansion that remains accurate
toward the classical limit. The resulting closed expression for the
phase correlator is inserted into the standard spherical-approximation
QRA without sacrificing the method’s flexibility with respect to lattice
geometry and dimensionality. The approach reproduces the shrinkage
of Mott lobes from $T=0$ up to $k_{B}T/U\simeq0.2$ in quantitative
agreement with theoretical predictions and with in-situ imaging experiments.
This \emph{finite-T QRA} thus supplies an analytic, computationally
light tool for strongly correlated lattice bosons and sets the stage
for amplitude-fluctuation upgrades required at higher temperatures.
\end{abstract}
\maketitle

\section{Introduction}

The past two decades have witnessed the transformation of ultracold‐atom
systems from delicate few‐body curiosities into fully fledged \emph{quantum
simulators} of strongly correlated lattice models. When a degenerate
gas of bosonic atoms is loaded into an optical standing-wave potential,
its low-energy physics is accurately captured by the Bose--Hubbard
Hamiltonian \citep{Fisher1989a,Jaksch1998}. By tuning $t/U$ (tunneling
amplitude vs. on-site interaction) one can drive a quantum phase transition
between a compressible superfluid (SF) and an incompressible Mott
insulator (MI). The earliest observation of this transition through
time-of-flight interference fringes already established optical lattices
as clean realizations of paradigmatic condensed-matter models \citep{Greiner2002}.
Today, quantum-gas microscopes provide single-site resolution of density
and coherence, allowing \emph{in situ} snapshots of individual atoms
and defects \citep{Bakr2010,Sherson2010}.

In solid-state materials the absolute lattice temperature (in kelvin)
is the natural metric because both the coupling constants and the
environment are fixed. In an optical lattice, by contrast, the interaction
scale $U$ is \emph{engineered} and typically lies in the tens of
nanokelvin range; moreover the lattice is raised almost adiabatically.
Under near-isentropic loading the entropy per particle, $s=S/N$,
is conserved to a good approximation, and the final temperature therefore
adjusts until the ratio $k_{B}T/U$ satisfies this entropy constraint
\citep{Bloch2008}. Reaching low \emph{dimensionless} temperatures,
e.g. $k_{B}T/U\lesssim0.05$, for bosonic MI order, thus hinges on
cooling the \emph{entropy reservoir} rather than the thermometer reading
in nanokelvin.

Precise thermometry in optical-lattice experiments remains notoriously
hard because the lattice suppresses conventional probes; as McKay
and DeMarco stress, reliable $T$ estimates almost always rely on
matching in-situ or parity-projected density profiles to quantum-Monte-Carlo
calculations rather than on direct fitting of time-of-flight images
\citep{McKay2011}. Even with that caveat, present bosonic systems
sit surprisingly “\emph{warm}’’ on the interaction scale set by $U$:
most laboratories report $k_{B}T/U\sim0.1-0.25$. Early single-site
microscopes already reached $k_{B}T/U=0.15$ in the first images of
the superfluid--Mott crossover \citep{Bakr2010}; subsequent Rb work
found $k_{B}T/U=0.09$ for $n=1$ (one atom per site) shell and $0.074$
for $n=2$ \citep{Sherson2010}. Similar values, $k_{B}T/U\approx0.1-0.21$,
were obtained for non-cooled Yb MI samples \citep{Miranda2017}, while
very light $^{7}\mathrm{Li}$ gave $k_{B}T/U\approx0.1$ (with a lower
bound of $0.083$ inferred from fits) \citep{Kwon2022}. The current
record belongs to Yang et al., who combined staggered-immersion cooling
with sublattice entropy removal to push a two-dimensional Rb array
down to $k_{B}T/U=0.046$ and a homogeneous entropy of only $1.9\times10^{-3}k_{B}$
per particle \citep{Yang2020}. Although the absolute temperatures
lie in the ten- to hundred-nano-kelvin range, they are comparable
to $U$, underscoring the challenge of cooling into the deep-quantum
regime where $k_{\mathrm{B}}T\ll U$. Thus, while genuine low-entropy
magnetism still lies ahead, today’s benchmarks unanimously place experimental
bosons at $k_{B}T/U\gtrsim0.05$. with most data clustered around
$0.1\lyxmathsym{–}0.2$5.\lyxdeleted{t.zaleski}{Thu Dec 18 09:47:07 2025}{ }

Capturing correlation physics in this entropy-limited regime is a
problematic theoretical challenge. Mean-field and local-density approximations
neglect critical phase fluctuations and therefore overestimate $T_{c}$,
$t_{c}/U$ and compressibility \citep{Dickerscheid2003,Lu2006}. Strong-coupling
perturbative-DMRG treatments are well-fitted for $k_{B}T/U\lesssim0.2$,
however require that tunnelling is small \citep{Plimak2004,DeMarco2005}.
On the other hand, controlled QMC simulations provide the gold standard
for equilibrium benchmarks \citep{Capogrosso2008,Mahmud2011,Capogrosso2007},
yet their computational cost scales significantly with system size
and inverse temperature, rendering them impractical for rapid scans
across lattice geometries or parameter space. To this end, an analytic,
geometry-agnostic approach that remains quantitatively reliable for
$k_{B}T\lesssim U$ could be very useful.

The quantum-rotor approach (QRA) meets part of this need: by integrating
out amplitude fluctuations and focusing on phase dynamics, it yields
closed-form expressions for the order parameter and phase boundary
that automatically incorporate lattice density of states and dimensionality
\citep{Polak2007}. At zero temperature it reproduces the universal
shape of Mott lobes and their critical exponents. Its scope, however,
extends far beyond static phase boundaries. The method quantitatively
reproduces the time-of-flight (TOF) momentum distributions seen in
clean lattices and, crucially, in the presence of synthetic magnetic
fields of arbitrary gauge\textbf{ }\citep{Zaleski2011TOF,ZaleskiPolak2011,PolakZaleski2013}.
By convolving the QRA phase propagator with a Bogoliubov treatment
of density modes, it yields single-particle spectral functions $A\left(\boldsymbol{k},\omega\right)$
across superfluid and Mott regimes, capturing the persistence (or
loss) of sharp coherence peaks and the transfer of spectral weight\textbf{
}\citep{Zaleski2012,Zaleski2012b}; or predicts dynamic structure
factor $S\left(\boldsymbol{k},\omega\right)$ directly measurable
by Bragg spectroscopy\textbf{ }\citep{Zaleski2017_StructFact}; or
allows to compute momentum-resolved longitudinal conductivity, predicting
linear low-$k$ behavior \citep{Grygiel2021}. However, its principal
weakness remains intrinsic: the phase correlator is still evaluated
with \emph{ground-state} statistics, so thermal winding numbers are
omitted. Although it still can be used to approximate the low-temperature
regime \citep{Zaleski2014,Zaleski2015}, the standard quantum rotor
approach may lead to oversharpened SF--MI crossover and underestimation
of the compressibility when $k_{B}T/U\gtrsim0.0022$.

In the present paper we construct a finite-temperature extension of
the quantum rotor model that remains quantitatively accurate up to
$k_{B}T/U\approx0.2$. Two complementary analytic expansions are developed:
(1) winding-number resummation; (2) auxiliary-field (high-$T$) expansion
with small parameter being $1/\beta U$. Both yield closed expressions
for the phase correlator that drop into the spherical approximation
without increasing computational effort. We show that the resulting
phase diagram and density profiles reproduce theoretical predictions
and experimental in-situ microscopy observations. The extended rotor
framework thus furnishes a versatile, computationally light tool for
strongly correlated lattice bosons, while its construction points
the way toward incorporating amplitude fluctuations and addressing
even higher-temperature, normal-fluid regimes.

\section{Quantum Rotor Approach to the Bose Hubbard model}

The Bose Hubbard model describes interacting bosons in a periodic
lattice with hopping between neighboring sites in the tight-binding
scheme \citep{Jaksch1998}. The Hamiltonian hence consists of an onsite
interaction term proportional to $U$, a hopping term dependent on
the hopping matrix $t_{ij}$, and a term proportional to the chemical
potential $\mu$, that controls the number of particles:
\begin{equation}
H=\frac{U}{2}\sum_{i}n_{i}\left(n_{i}-1\right)-\sum_{\left\langle i,j\right\rangle }t_{ij}a_{i}^{\dagger}a_{j}-\mu\sum_{i}n_{i},
\end{equation}
where $a_{i}^{\dagger}$, $a_{i}$ and $n_{i}=a_{i}^{\dagger}a_{i}$
are the usual bosonic creation, annihilation and number operators,
respectively. Throughout this paper we will make use of units in which
$U=1$ and define a shifted chemical potential $\bar{\mu}=\mu+1/2$.

Here, we employ the Quantum Rotor Approach (QRA) to study the Bose
Hubbard model \citep{Polak2007}. In the QRA, we make use of a path
integral formulation with a Wick rotation of the inverse temperature,
$\beta=1/k_{B}T$, into imaginary time, $\tau$, writing the partition
function as 
\begin{equation}
Z=\int\mathcal{D}\bar{a}\mathcal{D}a\:e^{-S\left[\bar{a},a\right]},
\end{equation}
with the action defined by 
\begin{equation}
S\left[\bar{a},a\right]=\sum_{i}\int_{0}^{\beta}d\tau\left[\bar{a}_{i}\left(\tau\right)\partial_{\tau}a_{i}\left(\tau\right)+H\left(\tau\right)\right].
\end{equation}
After getting rid of the quartic term with Hubbard-Stratonovich transformation
by means of an auxiliary field $V_{i}\left(\tau\right)=V_{i}+\partial_{\tau}\phi_{i}\left(\tau\right)$,
a local gauge transformation $a_{i}\left(\tau\right)\mapsto b_{i}\left(\tau\right)e^{i\phi_{i}\left(\tau\right)}$
introduces bosonic fields (amplitude and phase) in terms of which
the problem is reformulated underscoring the importance of the phase
coherence in formation of the superfluid state. It is important to
note that the phase fields $\phi_{i}\left(\tau\right)$ must be periodic
modulo $2\pi$ in imaginary time, $\phi_{i}\left(\tau+\beta\right)-\phi_{i}\left(\tau\right)=2\pi n_{i}$,
with $n_{i}\in\mathbb{Z}$ being the winding numbers.

The $b_{i}$ fields are then integrated out via a saddle point approximation
leading to an effective description in terms of the phase fields,
which defines a new phase order parameter $\psi=\left\langle e^{i\phi_{i}\left(\tau\right)}\right\rangle $.
The phase-only action is given by 
\begin{align}
S\left[\phi\right] & =\sum_{i}\int_{0}^{\beta}d\tau\left(\frac{1}{2}\dot{\phi}_{i}^{2}\left(\tau\right)+i\bar{\mu}\dot{\phi}_{i}\left(\tau\right)\right)\nonumber \\
 & +\sum_{\left\langle i,j\right\rangle }t_{ij}\left|b_{0}\right|^{2}\int_{0}^{\beta}d\tau\cos\left(\phi_{i}\left(\tau\right)-\phi_{j}\left(\tau\right)\right)
\end{align}
with $\dot{\phi}_{i}\left(\tau\right)=\partial_{\tau}\phi_{i}\left(\tau\right)$
and $\left|b_{0}\right|^{2}=\left(zt+\overline{\mu}\right)/U$ being
the static part of the $b_{i}\left(\tau\right)$ fields, i.e. $b_{0}=\left\langle b_{i}\left(\tau\right)\right\rangle $,
where $z$ is the coordination number and $t=t_{ij}$ (for hopping
between the nearest neighbors).

It is now convenient to introduce a unimodular field $z_{i}\left(\tau\right)$
via the resolution of the identity 
\begin{align}
1 & =\int\mathcal{D}\bar{z}\mathcal{D}z\prod\delta\left(z_{i}\left(\tau\right)-e^{i\phi_{i}\left(\tau\right)}\right)\nonumber \\
 & \times\delta\left(\bar{z}_{i}\left(\tau\right)-e^{-i\phi_{i}\left(\tau\right)}\right).
\end{align}
Next, we make use of the spherical approximation \citep{Berlin1952,Joyce1966},
which consists on relaxing the unimodularity of $z_{i}\left(\tau\right)$,
$\bar{z}_{i}\left(\tau\right)z_{i}\left(\tau\right)=1$, to an average
unimodularity, $\frac{1}{N}\sum_{i}\bar{z}_{i}\left(\tau\right)z_{i}\left(\tau\right)=1$,
by means of a Lagrange multiplier $\lambda$:
\begin{align}
 & \delta\left(\sum\bar{z}_{i}\left(\tau\right)z_{i}\left(\tau\right)-N\right)\\
 & \propto\int_{-i\infty}^{+i\infty}d\lambda\exp\left(N\lambda-\lambda\sum_{i}\bar{z}_{i}\left(\tau\right)z_{i}\left(\tau\right)\right).\nonumber 
\end{align}
After a Fourier transform into momentum and Matsubara frequency space,
we are left with a diagonal action:
\begin{align}
S\left[\bar{z},z\right] & =\frac{1}{\beta N}\sum_{\vec{k},\omega}\bar{z}\left(\vec{k},\omega\right)\nonumber \\
 & \left(\lambda-\left|b_{0}\right|^{2}T\left(\vec{k}\right)+\gamma^{-1}\left(\omega\right)\right)z\left(\vec{k},\omega\right),\label{eq:ActionZ}
\end{align}
where $T\left(\vec{k}\right)$ is the dispersion relation for the
lattice, and $\gamma\left(\omega\right)$ is the Fourier transform
of the phase correlator $\left\langle e^{i\phi_{i}\left(\tau\right)-i\phi_{i}\left(\tau'\right)}\right\rangle $,
which will be studied in detail in the next section.

The partition function is obtained after integrating out the $z_{i}\left(\tau\right)$
field reads:
\begin{equation}
Z=\int d\lambda\exp\left(N\beta\lambda+\sum_{\vec{k},m}\ln G\left(\vec{k},\omega_{m}\right)\right),\label{eq:partfunct}
\end{equation}
where 
\begin{equation}
G\left(\vec{k},\omega_{m}\right)=\text{\ensuremath{\left(\lambda-\left|b_{0}\right|^{2}T\left(\vec{k}\right)+\gamma^{-1}\left(\omega_{m}\right)\right)}}^{-1}\label{eq:prop}
\end{equation}
 is the propagator, $\omega_{m}=2\pi m/\beta$ is the Matsubara frequency
and $m\in\mathbb{Z}$. For a large number of lattice sites $N$, the
steepest descent method gives the value of the Lagrange multiplier
by fulfilling the following condition:
\begin{equation}
\frac{1}{N\beta}\sum_{\vec{k},m}G\left(\vec{k},\omega_{m}\right)=1.\label{eq:EquationForLambda}
\end{equation}
In\textbf{ }the superfluid region, the Lagrange multiplier sticks
to the value $\lambda_{SF}=\left|b_{0}\right|^{2}T\left(0\right)-\gamma^{-1}\left(0\right)$,
which signals the emergence of the ordered phase by the divergence
of the susceptibility. One thus obtains the equation for the order
parameter 
\begin{equation}
1-\psi^{2}=\frac{1}{N\beta}\sum_{\vec{k},m}G\left(\vec{k},\omega_{m}\right).
\end{equation}
It is useful to introduce now the density of states $\rho\left(x\right)=\frac{1}{N}\sum_{\vec{k}}\delta\left(x-T\left(\vec{k}\right)\right)$,
in terms of which the previous equation is written simply as: 
\begin{equation}
1-\psi^{2}=\int dx\rho\left(x\right)\frac{1}{\beta}\sum_{m}\frac{1}{\lambda-\left|b_{0}\right|^{2}x+\gamma^{-1}\left(\omega_{m}\right)}.\label{eq:EquationOfState}
\end{equation}
In the present paper, we are only making use of the density of states
corresponding to the simple cubic lattice. However, it is straightforward
to extend the results to other geometries simply by choosing $\rho\left(x\right)$
appropriately \citep{Zaleski2015,Polak2009,Zaleski2010,Zaleski2010b}.

\section{Computation of the correlator}

The finite temperature effects on the QRA are mainly encoded in the
phase correlator, which can be written as a sum over winding numbers:
\begin{align}
\gamma\left(\omega_{m}\right) & =\frac{1}{Z_{0}}\sum_{n=-\infty}^{\infty}e^{-{\displaystyle \beta\left(n-v\left(\mu\right)\right)^{2}/2}}\nonumber \\
 & \times\frac{1}{1/4-\left(n-v\left(\mu\right)-i\omega_{m}\right)^{2}},\label{eq:corr}
\end{align}
with the free rotors partition function $Z_{0}=\sum_{n=-\infty}^{\infty}e^{-{\displaystyle \beta\left(n-v\left(\mu\right)\right)^{2}/2}}$
and the function $v\left(x\right)=x-\left\lfloor x\right\rfloor -1/2$,
where $\left\lfloor x\right\rfloor $ denotes the floor function,
which gives the greatest integer less than or equal to $x$ (the periodicity
of $v(x)$ is a direct consequence of the periodicity of the phase
variables at each lattice site).\lyxdeleted{t.zaleski}{Thu Dec 18 09:47:07 2025}{ }

It should be noted that in the equation of state in Eq. (\ref{eq:EquationOfState})
the inverse of the correlator is used, which requires it to be in
closed and simple form to be able to perform the summation of the
Matsubara frequencies. In the original QRA method in Ref. \citep{Polak2007}
the phase correlator was evaluated in the zero-temperature limit $\beta\rightarrow\infty$,
in which the summation in Eq. (\ref{eq:corr}) simplified to just
one leading term with the winding number $n_{min}$, for which the
$n-v\left(\mu\right)$ was minimal, i.e.:
\begin{equation}
\gamma^{-1}\left(\omega_{m}\right)=1/4-\left(n_{min}-v\left(\mu\right)-i\omega_{m}\right)^{2}.
\end{equation}

For the general form of the correlator, the sum of the series in Eq.
(\ref{eq:corr}) does not exist in a closed form. To go beyond the
zero-temperature limit, in the following we present two different
approximations which allow to obtain closed form of the inverse correlator
on the Matsubara frequencies and perform the summation in Eq. (\ref{eq:EquationOfState}).

\subsection{Winding number expansion}

The simplest approximation, which constitutes a low temperature expansion,
is based on taking into account more than one winding number in the
Eq. (\ref{eq:corr}). These winding numbers should provide the leading
terms of the sum, which again, occurs when the values of $n-v\left(\mu\right)$
are the lowest. A compact way of writing this is by defining $\tilde{v}\left(\mu\right)=v\left(\mu\right)+\left\lfloor 2\left\{ \mu\right\} \right\rfloor \left(1-2\left\{ \mu\right\} \right)$,
which maps the fractional part to the interval $\left[0,1/2\right]$
symmetrically around $1/2$, and making use of the symmetry (antisymmetry)
of the real (imaginary) part of the correlator with respect to this
mapping. In the simplest case, we take the two leading winding numbers:
\begin{align}
\gamma\left(\omega_{m}\right) & \approx\frac{1}{Z_{0}}\left\{ \frac{e^{-{\displaystyle \beta\tilde{v}^{2}\left(\mu\right)/2}}}{1/4-\left(\tilde{v}\left(\mu\right)-\left(-1\right)^{\left\lfloor 2\left\{ \mu\right\} \right\rfloor }i\omega_{m}\right)^{2}}\right.\nonumber \\
 & \left.+\frac{e^{-{\displaystyle \beta\left(\tilde{v}\left(\mu\right)+1\right)^{2}/2}}}{1/4-\left(\tilde{v}\left(\mu\right)+1-\left(-1\right)^{\left\lfloor 2\left\{ \mu\right\} \right\rfloor }i\omega_{m}\right)^{2}}\right\} \label{eq:corr_2windingnbrs}
\end{align}

\begin{figure}
\begin{centering}
\includegraphics[scale=0.47]{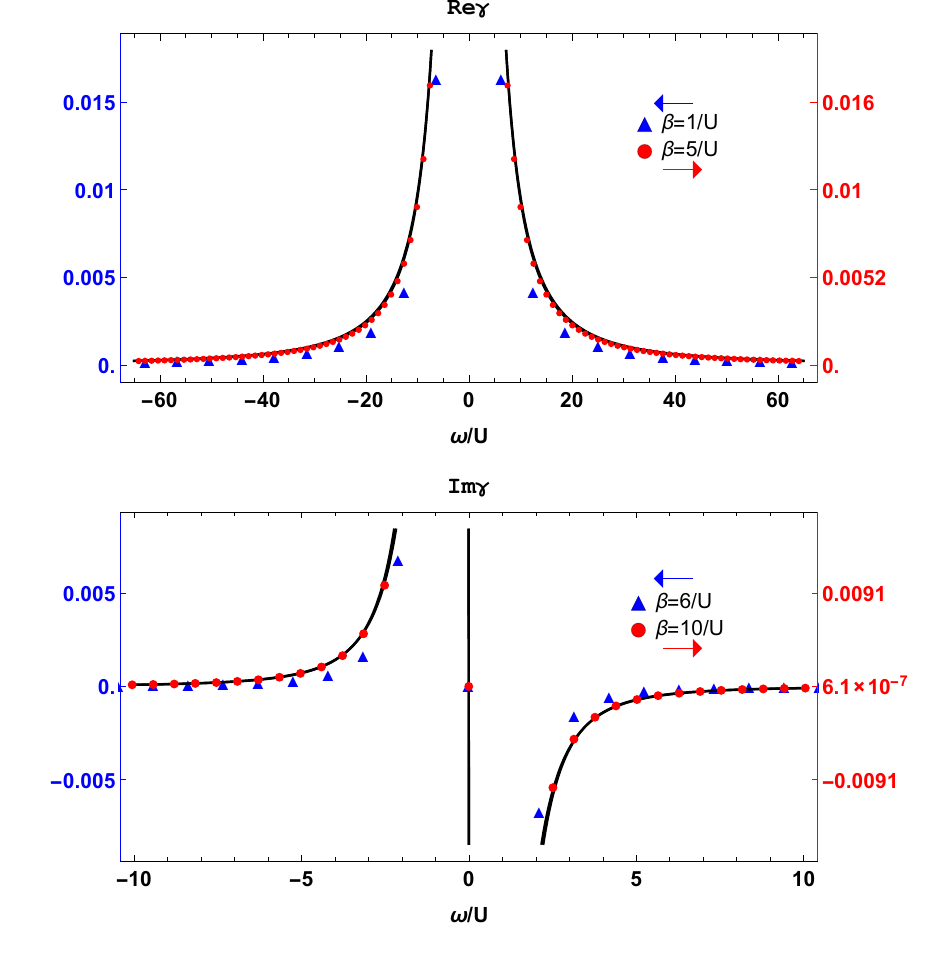}
\par\end{centering}
\caption{Real (top) and imaginary (bottom) parts of the phase correlator: numerical
result (solid line) vs. two-term winding-number approximation (colored
symbols at Matsubara frequencies) for $\mu/U=0.4$. Each panel shows
two temperatures: a lower one with good agreement (right vertical
axis) and a higher one where deviations become noticeable (left vertical
axis).}\label{fig:2terms}
\end{figure}

The higher the temperature, the more winding numbers need to be taken
into account in this approximation. However, just two winding numbers
appear to provide a very good approximation of the original correlator
up to temperatures of the order $\beta\sim1/U$ ($k_{B}T\sim U$),
where the difference starts to be considerable. This can be checked
by comparing the values provided by Eq. (\ref{eq:corr_2windingnbrs})
with numerical evaluation of Eq. (\ref{eq:corr}) for large range
of $n$ (in our case, $n=-300,\dots,300$). The results are shown
in Figure \ref{fig:2terms}. The temperatures shown for the real (Re)
and imaginary (Im) parts differ. Each panel in Fig. \ref{fig:2terms}
includes one temperature where the approximation agrees with the numerical
result and one where deviations become noticeable, to indicate the
range of validity.

It can be noted that the accuracy of the two-winding number expansion
could be improved by generalizing the forms of numerators in the Eq.
(\ref{eq:corr_2windingnbrs}) to read:\lyxdeleted{t.zaleski}{Thu Dec 18 09:47:07 2025}{ }

\emph{
\begin{eqnarray}
 &  & \gamma\left(\omega_{m}\right)=\frac{a\left(\beta,\mu\right)}{1/4-\left(\tilde{v}\left(\mu\right)-\left(-1\right)^{\left\lfloor 2\left\{ \mu\right\} \right\rfloor }i\omega_{m}\right)^{2}}\nonumber \\
 &  & +\frac{b\left(\beta,\mu\right)}{1/4-\left(\tilde{v}\left(\mu\right)+1-\left(-1\right)^{\left\lfloor 2\left\{ \mu\right\} \right\rfloor }i\omega_{m}\right)^{2}}\label{eq:corr_2windingnbr_crctd}
\end{eqnarray}
}with $a$ and $b$ being interpolating functions and fitting their
values so the formula in Eq. (\ref{eq:corr_2windingnbr_crctd}) reproduces
numerically calculated values of the correlator better than the one
in Eq. (\ref{eq:corr_2windingnbrs}). However, this does not seem
to significantly broaden the approximation's usable temperature range
while simultaneously rendering its justification and controllability
more uncertain.

The proposed approximation leads to rational expressions for the inverse
correlator, which can be decomposed into partial fractions. As a result,
the summation of the propagator over Matsubara frequencies in Eq.
(\ref{eq:EquationOfState}) can be performed analytically. This, however,
puts a practical limitation on including more winding numbers, as
the process is significantly more complicated the higher is the order
of the polynomials involved. As a result, a different approach to
taking more terms in the winding number expansion would be desirable
for a broader temperature range of the approximated correlator.

\subsection{Expansion on an auxiliary variable}

To obtain a broad temperature approximation for the correlator, we
define the following function $S(x)$ of an auxiliary variable $x$:
\begin{equation}
S\left(x\right)=\sum_{n}{\displaystyle \frac{e^{-\left(an^{2}x+bnx+cn\right)}}{{\displaystyle d+n^{2}+\frac{b}{a}n}}}\label{eq:sx_aux}
\end{equation}
with complex parameters $a$, $b$, $c$, and $d$ (being functions
of $\mu$, $\beta$ and $\omega_{m}$). Taking the first derivative
of $S(x)$, one arrives at a differential equation:
\begin{equation}
S'\left(x\right)=-a\sum_{n}{\displaystyle e^{-\left(an^{2}x+bnx+cn\right)}}+adS(x),\label{eq:diffeq}
\end{equation}
which cannot be solved in closed form. However, we note that the phase
correlator in Eq. (\ref{eq:corr}) can be written in terms of $S\left(x\right)$
in the limit $x\rightarrow1$ as:
\begin{equation}
\gamma\left(\omega_{m}\right)=\frac{A}{Z_{0}}\lim_{x\rightarrow1}S(x)
\end{equation}
by choosing:
\begin{align}
a=\beta/2\qquad & b=-\beta\left(v\left(\mu\right)+i\omega_{m}\right)\nonumber \\
c=0\:\left(\mathrm{bosons}\right)\qquad & d=\left(v\left(\mu\right)+i\omega_{m}\right)^{2}-1/4,
\end{align}
with $A=-e^{\beta v^{2}\left(\mu\right)}$ and noting the discrete
nature of the Matsubara frequencies, for which $e^{\pm\beta i\omega_{m}}=1$.
Since only the point $x=1$ is important in Eq. (\ref{eq:diffeq}),
its first term can be expanded in a power series around this point,
giving rise to a polynomial, for which the differential equation can
be analytically solved at any finite order of the expansion. At order
$N$, the approximate solution to $S(x)$ is
\begin{align}
 & S_{N}(x)=e^{adx}\left(C_{N}-aZ'e^{-ad}\sum_{j=0}^{N}\left(-1\right)^{j}\right.\nonumber \\
 & \times\left\langle \left(an^{2}+bn\right)^{j}\right\rangle '\frac{1}{\left(ad\right)^{1+j}}\nonumber \\
 & \times\left.\left\{ 1-e^{-ad\left(x-1\right)}\sum_{k=0}^{j}\frac{1}{k!}\left[ad\left(x-1\right)\right]^{k}\right\} \right),
\end{align}
where $C_{N}$ depends on the boundary conditions and $\left\langle \cdot\right\rangle '$
and $Z'$ are the expectation values and the partition function, respectively,
for the distribution ${\displaystyle e^{-\left(an^{2}+bn+cn\right)}},n\in\mathbb{Z}$.
This leads to the approximation for the correlator (at $x\rightarrow1$):
\begin{equation}
\gamma\left(\omega_{m}\right)\approx\frac{A}{Z}C_{N}e^{ad}.
\end{equation}
The constant $C_{N}$ can be determined from the value of $S\left(0\right)$,
which can be expressed in terms of hypergeometric functions. For the
problem at hand $Z'=e^{-\beta v^{2}\left(\mu\right)}Z_{0}$, which
leads to the final expression for the correlator at $N$-th order
of expansion given by
\begin{align}
\gamma\left(\omega\right) & \approx\sum_{j=0}^{N}\left\langle \left(n^{2}+\frac{b}{a}n\right)^{j}\right\rangle _{Z_{0}}\frac{1}{\left(-d\right)^{1+j}}\nonumber \\
 & \times\left[1-e^{\frac{\beta}{2}d}\sum_{k=0}^{j}\frac{1}{k!}\left(\frac{\beta}{2}\right)^{k}\left(-d\right)^{k}\right],\label{eq:auxexp}
\end{align}
which can be shown to converge to the correlator in Eq. (\ref{eq:corr}).

The fact that the sum in Eq.\,(\ref{eq:sx_aux}) runs over all integers
$n$ means that any shift $n\rightarrow n+k$ (with $k\in\mathbb{Z}$)
leaves the correlator unchanged. In other words, since shifting the
summation index does not alter $\gamma\left(\omega_{m}\right)$, one
may think there is a freedom in choosing the parameters $a$, $b$,
$c$ and $d$. However, to preserve the even symmetry of the real
part and the odd symmetry of the imaginary part under $\omega_{m}\rightarrow-\omega_{m}$,
these parameters must be fix exactly as above. Any other assignment,
even if equivalent only up to an integer shift, would break those
symmetries, forcing a “re-symmetrization” analogous to what was done
for the winding-number expansion (there, by defining the symmetric
$\tilde{v}\left(\mu\right)$ mapping).\lyxdeleted{t.zaleski}{Thu Dec 18 09:47:07 2025}{ }

\begin{figure}
\begin{centering}
\includegraphics[scale=0.5]{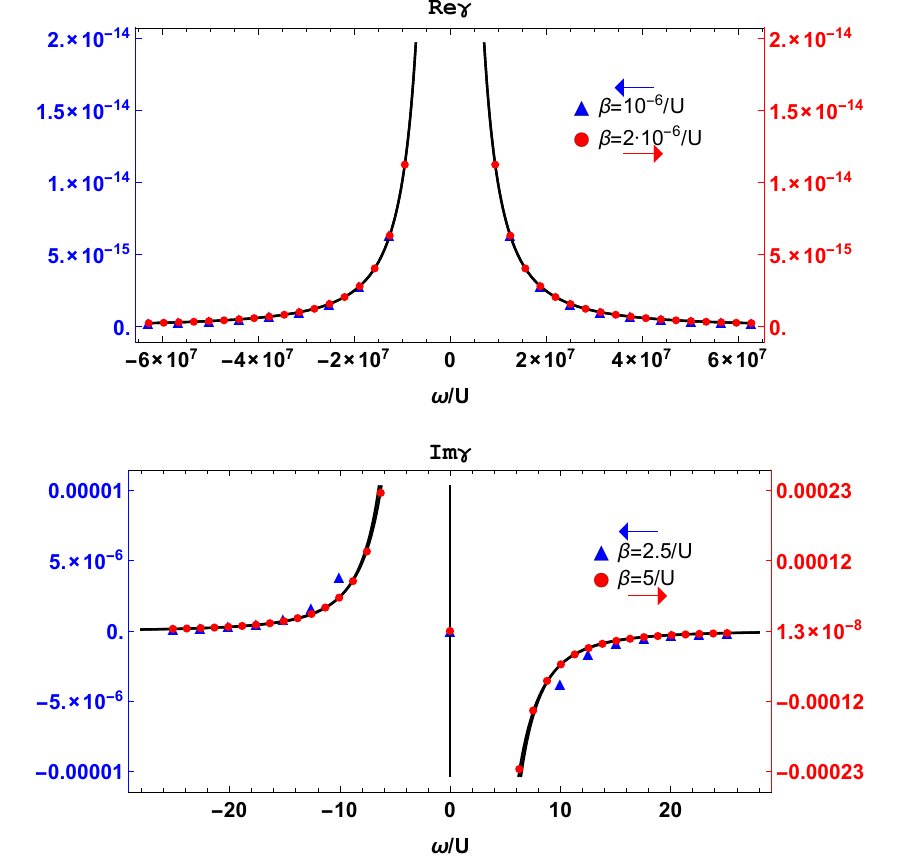}
\par\end{centering}
\caption{Real (top) and imaginary (bottom) parts of the phase correlator: numerical
result (solid line) vs. two-term auxiliary-variable approximation
(colored symbols at Matsubara frequencies) for $\mu/U=0.4$. Each
panel shows two temperatures. In the top panel, both temperatures
display excellent agreement, indicating that the approximation remains
accurate even at high temperature. In the bottom panel, the lower
temperature still shows good agreement (right vertical axis), while
the higher temperature illustrates the onset of deviation (left vertical
axis). }\label{fig:auxvar}
\end{figure}

\begin{figure}
\begin{centering}
\includegraphics[scale=0.5]{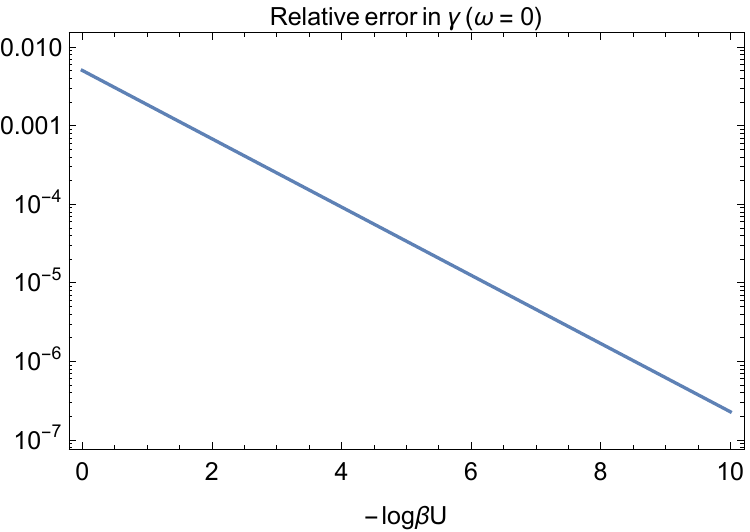}
\par\end{centering}
\caption{Decrease of the relative error in the estimation of $\gamma\left(\omega_{m}=0\right)$
in Eq. (\ref{eq:omega0}) with the numerically evaluated value of
Eq. (\ref{eq:corr}) for summation over $n=-300,\dots,300$ while
the temperature is being increased.}\label{fig:err}
\end{figure}

Figure \ref{fig:auxvar} shows the real and imaginary parts of the
expansion in Eq. (\ref{eq:auxexp}) at order $N=1$. The real part
is accurately reproduced at all temperatures except at $\omega_{m}=0$
point, at which it is only accurate for temperatures below \foreignlanguage{polish}{$\beta\sim30/U$}.
\foreignlanguage{polish}{This zero-frequency point is important because
strongly affects the phase boundary and particle density via the Lagrange
multiplier. For this reason, we handle it more carefully below, with
the corresponding} results summarized in Fig. \ref{fig:err}.\lyxdeleted{t.zaleski}{Thu Dec 18 09:47:07 2025}{ }

The imaginary part remains accurate only for $\beta\gtrsim2.5/U$;
two temperatures are shown in Fig. \ref{fig:auxvar} to illustrate
where the approximation begins to deviate from the numerical result.
However, since the imaginary part becomes small at high temperatures
in both the numerical result and the approximation, the auxiliary-variable
expansion remains practically useful over a broader temperature range
than the winding-number expansion at the same order.

The slower convergence of the imaginary part is due to the expansion
of the $e^{i\beta\omega_{m}nx}$ factor. This can be improved by rewriting
it as $e^{i2\pi\left\{ nmx\right\} }$ for each Matsubara frequency
$\omega_{m}=2\pi m/\beta$ and then performing the Taylor expansion.
However, this leads to an increase of the complexity of the resulting
expressions while not meaningfully improving the quality of the approximation.

Regarding the failure of this expansion to retrieve the correct value
of the real part of the correlator at $\omega_{m}=0$ at low orders
of expansion, one could simply use the value from the numerical calculation
of the correlator at this point. Nevertheless, it is possible to obtain
a closed form expression in terms of special functions by rewriting
the $\sum{\displaystyle e^{-\left(an^{2}x+bnx+cn\right)}}$ term in
Eq. (\ref{eq:diffeq}) in terms of the Jacobi theta function $\vartheta_{3}\left(z,q\right)$:
\begin{align}
 & \sum{\displaystyle e^{-\left(an^{2}x+bnx+cn\right)}}=e^{{\displaystyle \frac{\left(c+bx\right)^{2}}{4ax}}}\sqrt{\frac{\pi}{ax}}\nonumber \\
 & \times\vartheta_{3}\left(\frac{\pi\left(c+bx\right)}{2ax},e^{-{\displaystyle \frac{\pi^{2}}{ax}}}\right)
\end{align}
and taking the series expansion only in $\vartheta_{3}$, leading
to
\begin{align}
\gamma\left(\omega_{m}\right)\approx & \frac{2\pi}{Z}e^{-\frac{\beta}{2}d}\left(\mathrm{Im}\left(\mathrm{erf}\left(i\sqrt{\frac{\beta}{8}}\right)\right)\right.\nonumber \\
 & -2\cos\left(2\pi\left(v\left(\mu\right)+i\omega_{m}\right)\right)\nonumber \\
 & \times\left.\mathrm{Im}\left(\mathrm{erf}\left(\pi\sqrt{\frac{2}{\beta}}-i\sqrt{\frac{\beta}{8}}\right)\right)\right),\label{eq:omega0}
\end{align}
where $\mathrm{erf}\left(z\right)=\frac{2}{\sqrt{\pi}}\int_{0}^{z}e^{-t^{2}}\,dt$
is the complex error function. The expression in Eq. (\ref{eq:omega0})
not only works near $\omega_{m}=0$ but also gets better the higher
the temperature is, as shown in Figure \ref{fig:err}.

In order to justify the success of this expansion in reproducing the
correct values of the correlator at arbitrarily high temperatures,
it is convenient to rewrite the phase correlator in Eq. (\ref{eq:corr})
as 
\begin{align}
\gamma\left(\omega_{m}\right) & =\frac{\beta}{2}\int_{-1}^{0}d{\displaystyle te^{-{\displaystyle \beta td\left(\omega_{m}\right)/2}}}\nonumber \\
 & \times\left\langle e^{-{\displaystyle \beta t\left(n^{2}/2-nv\left(\mu\right)-i\omega_{m}n\right)}}\right\rangle _{Z_{0}}.
\end{align}

Both schemes in effect only expand the expectation‐value part inside
the integral: winding numbers expand around $\beta\rightarrow\infty$
(low temperatures) and the auxiliary variable is equivalent to expansion
around $\beta\rightarrow0$ (high temperatures). Crucially, both leave
the overall factor $e^{-{\displaystyle \beta td\left(\omega_{m}\right)/2}}$
untouched, so even the high-$T$ (auxiliary-variable) expansion continues
to capture the dominant low-temperature behavior correctly.

\section{Bose Hubbard model at finite temperatures}\label{sec:Bose-Hubbard-model}

Making use of the approximations for the phase correlator discussed
in the previous section, it is possible to obtain analytical expressions
for the critical line and the particle density at non-zero temperatures.
The focus will be on temperatures corresponding to $\beta\gtrsim5/U$,
and hence, the appropriate and sufficient choice for the approximation
of the phase correlator is a two-winding number expansion, which also
facilitates analytical calculations. For instance, after performing
the Matsubara summation in Eq. (\ref{eq:EquationOfState}), the order
parameter is given by
\begin{align}
 & 1-\psi^{2}=-\frac{e^{-\beta\tilde{v}^{2}\left(\mu\right)/2}+e^{-\beta\left(\tilde{v}\left(\mu\right)+1\right)^{2}/2}}{2Z_{0}}\nonumber \\
 & \times\int dx\rho\left(x\right)\sum_{j=1}^{3}\coth\beta r_{j}(x)/2\nonumber \\
 & \times\frac{\tilde{v}\left(\mu\right)+r_{j}(x)+\frac{1}{2}\frac{3e^{-\beta\tilde{v}^{2}\left(\mu\right)/2}-e^{-\beta\left(\tilde{v}\left(\mu\right)+1\right)^{2}/2}}{e^{-\beta\tilde{v}^{2}\left(\mu\right)/2}+e^{-\beta\left(\tilde{v}\left(\mu\right)+1\right)^{2}/2}}}{\prod_{i\neq j}\left(r_{j}(x)-r_{i}(x)\right)},
\end{align}
where $z=r_{j}(x)$ are the zeros of the now rational function $D\left(z,x\right)=1+c(x)\gamma\left(z\right)$,
which results from the denominator of the propagator in Eq. (\ref{eq:prop})
with factored out $\gamma^{-1}\left(z\right)$, namely $\gamma^{-1}\left(z\right)$$\left[1+\left(\lambda-\left|b_{0}\right|^{2}x\right)\gamma\left(z\right)\right]$,
which leads to $c\left(x\right)=\lambda-\left|b_{0}\right|^{2}x$.
In general, at the $N$-th order of the winding number expansion:
\begin{align}
1-\psi^{2} & =-\frac{\sum_{j=2-N}^{N-1}e^{-\beta\left(\tilde{v}\left(\mu\right)+j\right)/2}}{2Z_{0}}\nonumber \\
 & \times\int dx\rho\left(x\right)\coth\beta r_{j}(x)/2\nonumber \\
 & \times\sum_{j=1}^{N+1}\frac{\prod_{i=1}^{N-1}\left(r_{j}(x)-q_{i}\right)}{\prod_{i\neq j}\left(r_{j}(x)-r_{i}(x)\right)},
\end{align}
with $z=q_{i}$ being the zeros of the extension of the phase correlator
to the whole complex plane, $\gamma\left(q_{i}\right)=0$.

\begin{figure}
\begin{centering}
\includegraphics[scale=0.75]{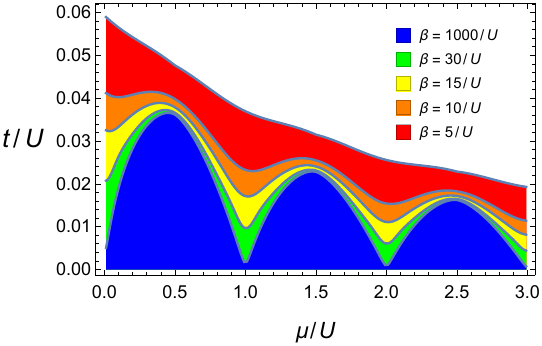}
\par\end{centering}
\caption{Phase diagram at different temperatures.}\label{fig:diagr}
\end{figure}

We can then easily obtain the phase diagram, i.e. the lines, where
the order parameter $\psi$ vanishes. Figure \ref{fig:diagr} shows
the phase diagram in the $\mu-t$ parameter space at different temperatures.
It shows how the low temperature lobe structure of the Mott insulator
progressively fades away when increasing the temperature, as well
as an increase on the hopping strength needed to reach a long range
phase coherent state in order to compensate for the increasing temperature
fluctuations.

\begin{figure}
\includegraphics[scale=0.75]{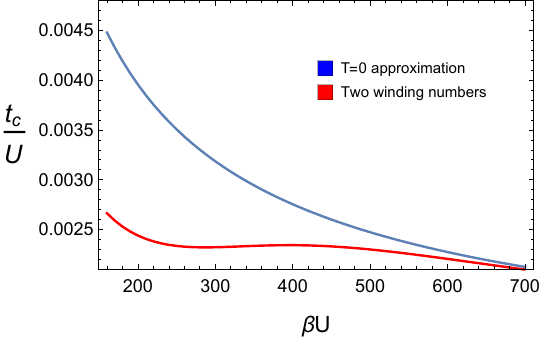}\caption{\foreignlanguage{american}{Comparison between the critical hopping parameter at $\mu/U=1$ for
the zero temperature approximation and the winding number expansion
with two terms.}}\label{fig:T0_Tn0_comparison}
\end{figure}

We can also use the above results to assess the region of validity
of the original QRA approach with the zero-temperature correlator.
In Figure \ref{fig:T0_Tn0_comparison} we have shown a comparison
between the critical hopping parameter at $\mu/U=1$ of the QRA with
the zero-temperature correlator and the winding number expansion.
Although, the deviation occurs even for very low temperatures, it
becomes visibly significant for $k_{B}T/U\approx0.002$ ($\beta U\approx460$,
where the deviation exceeds 10\%). \foreignlanguage{american}{The
zero temperature approximation leads not only quantitative differences
with respect to the finite temperature corrections, but also qualitative,
such as a sharpening of the phase diagram around integer values.}

Next, we obtain an analytical expression for the particle density
$n=df/d\mu$ (with $f=-\left(\beta N\right)^{-1}\ln Z$ being the
free energy per particle) in terms of the order parameter, the zeros
$\left(r_{j}\right)$ and the poles $\left(p_{j}\right)$ of $D\left(z,x\right)=1+c(x)\gamma\left(z\right)$:
\begin{align}
n & =\left|\psi\right|^{2}\partial_{\mu}\lambda+\bar{\mu}+\frac{1}{\beta}\partial_{\mu}\ln Z_{0}\nonumber \\
 & +\frac{1}{\beta}\partial_{\mu}\int dx\rho\left(x\right)\ln\left(-\frac{\prod\sinh\beta r_{j}(x)/2}{\prod\sinh\beta p_{j}(x)/2}\right)\nonumber \\
 & +\frac{1}{2}\partial_{\mu}\int dx\rho\left(x\right)\sum\left(s\left(r_{j}\right)r_{j}-s\left(p_{j}\right)p_{j}\right),
\end{align}
 where $s(z)$ is a sign function, 
\begin{equation}
s\left(z\right)=\left\{ \begin{array}{cc}
1 & \mathrm{Re}z\leq0\\
-1 & \mathrm{Re}z>0
\end{array}\right..
\end{equation}
Note that since the roots come in conjugate pairs and the poles are
real, it is irrelevant in which of the two cases $\mathrm{Re}z=0$
is included.

\begin{figure}
\begin{centering}
\includegraphics[scale=0.75]{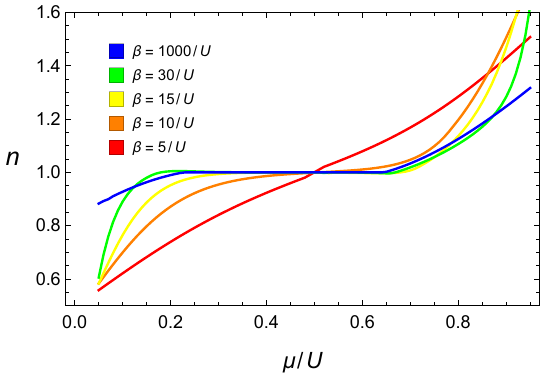}
\par\end{centering}
\caption{Particle density at different temperatures for hopping strength $t/U=0.03$.}\label{fig:dens}
\end{figure}

Figure \ref{fig:dens} shows the particle density at different temperatures
in terms of the chemical potential for a fixed hopping strength $t/U=0.03$.
We can clearly observe the fixed integer density in the Mott insulator
at low temperature, while the increase of temperature makes the density
evolve towards a linear relation. The results corroborate theoretical
predictions \citep{Gerbier2007,Mahmud2011} and experimental observations
\citep{Bakr2010,Sherson2010}. The incompressible Mott-insulating
phase begins to melt around $k_{B}T/U\approx0.1$, where the density
plateau in $\mu$ contracts; by $k_{B}T/U\approx0.2$ the Mott lobes
are washed out and $n(\mu)$ is almost linear.\lyxdeleted{t.zaleski}{Thu Dec 18 09:47:07 2025}{ }

The particle density can be also written in a more general way to
accommodate approximations of the correlator $\gamma\left(\omega_{m}\right)$
in Eq. (\ref{eq:corr}), which cannot be expressed as rational expressions
(this applies also when the correlator is evaluated purely numerically).
In such case, when $D\left(z,x\right)$ is no longer rational function:
\begin{align}
n & =\left|\psi\right|^{2}\partial_{\mu}\lambda+\bar{\mu}+\frac{1}{\beta}\partial_{\mu}\ln Z_{0}\nonumber \\
 & +\partial_{\mu}\int dx\rho\left(x\right)\frac{1}{\beta}\sum_{\omega}\ln\left(D\left(i\omega,x\right)\right)\nonumber \\
 & +\frac{1}{4\pi i}\partial_{\mu}\int dx\rho\left(x\right)\left(\oint_{\varGamma_{-}}dz\frac{zD'\left(z,x\right)}{D\left(z,x\right)}\right.\nonumber \\
 & \left.-\oint_{\varGamma_{+}}dz\frac{zD'\left(z,x\right)}{D\left(z,x\right)}\right)
\end{align}
with $\varGamma_{-}$ $\left(\varGamma_{+}\right)$ being a counterclockwise
contour in the left (right) half-plane containing all roots and poles
in that half, and $D'\left(z,x\right)$ denotes the derivative with
respect to the complex variable $z$.\lyxdeleted{t.zaleski}{Thu Dec 18 09:47:07 2025}{ }

The function $D\left(z,x\right)$ has the same number of poles and
zeros in each half-plane, so the branch cuts of $\ln D\left(z,x\right)$
can be chosen by forming pairs in between them, allowing the creation
of contours on each half-plane that do not cross any branch cuts and
still contain all zeros and poles of $D\left(z,x\right)$. Thus, the
last term can be rewritten as an integral of $\ln D\left(z,x\right)$,
which choosing large semicircles as contours, gives:
\begin{align}
n & =\left|\psi\right|^{2}\partial_{\mu}\lambda+\bar{\mu}+\frac{1}{\beta}\partial_{\mu}\ln Z_{0}\nonumber \\
 & +\partial_{\mu}\int dx\rho\left(x\right)\left(\frac{1}{\beta}\sum_{m}\ln\left(D\left(i\omega_{m},x\right)\right)\right.\nonumber \\
 & \left.-\frac{1}{2\pi}\int_{-\infty}^{\infty}d\omega\ln\left(D\left(i\omega_{m},x\right)\right)\right).
\end{align}
\lyxdeleted{t.zaleski}{Thu Dec 18 09:47:07 2025}{ }

The last term is now seen as the difference between the finite temperature
Matsubara sum of $D\left(z,x\right)$ and the sum at zero temperature.
In cases where the Matsubara sum cannot be performed analytically
and/or is computationally expensive (specially taking into account
that it will have to be averaged with respect to the density of states),
one can make use of the Euler-Maclaurin formula, which only requires
information about the derivatives of $\ln D\left(z,x\right)$ at $z=0$,
and take advantage of the rapid decrease of $B_{k}/k!$, with $B_{k}$
the $k$-th Bernoulli number:
\begin{align}
n & =\left|\psi\right|^{2}\partial_{\mu}\lambda+\bar{\mu}+\frac{1}{\beta}\partial_{\mu}\ln Z_{0}\nonumber \\
 & +\frac{1}{2\beta}\partial_{\mu}\int dx\rho\left(x\right)D\left(0,x\right)\nonumber \\
 & -\frac{1}{\beta}\sum_{k=2}^{\infty}{\displaystyle \frac{B_{k}}{k!}}\partial_{\mu}\int dx\rho\left(x\right)D^{(k-1)}\left(0,x\right),
\end{align}
 where $D^{(k)}\left(0,x\right)$ is the $k$-th derivative of $D\left(z,x\right)$
with respect to $z$ evaluated at $z=0$ and the sum can be truncated
at finite order.

\section{Conclusions}

We have extended the quantum-rotor approach (QRA) to correctly include
thermal fluctuations, providing a practical analytic tool for studying
the Bose-Hubbard model at finite temperature. The zero-temperature
phase correlator of Ref. \citep{Polak2007} is now replaced by two
complementary schemes: (i) a systematic expansion that keeps higher
winding numbers, and (ii) an auxiliary-variable representation that
rewrites functions of the form $f\left(\omega\right)=\sum_{n}f_{n}\left(\omega\right)$
via a Feynman-type integral identity. Both schemes yield closed-form
expressions for the correlator as a function of Matsubara frequency,
allowing seamless integration into the QRA framework. While the winding-number
expansion already covers the temperature range relevant to optical-lattice
experiments, the auxiliary-variable method is more general and should
prove valuable in related problems that require even higher temperature.

With this extension we have computed the finite-temperature phase
diagram of the Bose-Hubbard model together with the particle-density
profile. The results corroborate theoretical predictions \citep{Gerbier2007,Mahmud2011}
and experimental observations \citep{Bakr2010,Sherson2010}, namely
they correctly predict the melting of the incompressible Mott-insulating
phase at temperatures around $k_{B}T/U\approx0.1$, and loss of the
Mott lobes by $k_{B}T/U\approx0.2$. We have further shown that the
original QRA already breaks down at $k_{B}T/U\approx0.0022$, whereas
the present formulation remains easily accurate for $k_{B}T\gtrsim U$
-- well beyond the regime required for current experiments.

Beyond the observables presented here, the finite-T QRA retains the
versatility of its $T=0$ parent. It can therefore be applied directly
to momentum-resolved spectral functions, structure factors, and transport
kernels also in synthetic gauge fields at experimentally relevant
temperatures. The formalism also prepares the ground for incorporating
amplitude (Higgs) fluctuations, which are essential above $k_{B}T/U\approx0.2$.

In summary, this work closes the temperature gap that has separated
analytic QRA predictions from ultracold-atom data, providing a transparent
and computationally efficient framework for exploring strongly correlated
lattice bosons in the finite-entropy regimes accessible to experiment.

\section*{Acknowledgements}

\emph{Acknowledgments}. This research was funded in whole or in part
by National Science Centre, Poland within PRELUDIUM BIS-2 UMO-2020/39/O/ST3/01148
project.

\emph{Data availability}. The data that support the findings of this
paper are openly available \citep{zenododata}.

\bibliographystyle{apsrev4-2-local}
\bibliography{Bibliography_finitetemp}

\end{document}